\documentstyle[12pt,epsfig]{article}
\hoffset = - 0.6 truecm
\def\beq{\begin{equation}}   \def\eeq{\end{equation}}
\def\bea{\begin{eqnarray}}  \def\eea{\end{eqnarray}} 
\def\noi{\noindent} \def\beeq{\begin{eqnarray}}
\def\eeeq{\end{eqnarray}}
\def\lsim{\raise0.3ex\hbox{$<$\kern-0.75em\raise-1.1ex\hbox{$\sim$}}}
\def\gsim{\raise0.3ex\hbox{$>$\kern-0.75em\raise-1.1ex\hbox{$\sim$}}}

\begin{document} \begin{center}
\vbox to 1 truecm {}
{\large \bf J/$\psi$ suppression as a
function of the energy of the zero} \par \vskip 3 truemm
{\large \bf  degree calorimeter : can it
discriminate between} \par \vskip 3 truemm
{\large \bf deconfining and comovers interaction models~?}\vskip 1
truecm {\bf A. Capella and D. Sousa} \vskip 3 truemm

{\it Laboratoire de Physique Th\'eorique, UMR 8627 CNRS,\\ Universit\'e
Paris XI, B\^atiment 210, 91405 Orsay Cedex, France} \end{center}
\vskip 2 truecm

\begin{abstract}  The ratio of $J/\psi$ over Drell-Yan cross-sections as a
function of the energy of the zero degree calorimeter in $Pb$ $Pb$
collisions has been computed in a comovers interaction model and
compared with the results of a deconfining model. The predictions of
the two models are different both for peripheral events and for very
central ones. These differences are analyzed and the results of the
models confronted with
available data -- not only for $Pb$ $Pb$ but also for $pA$ and $SU$
collisions. \end{abstract}
\vskip 3 truecm
%\noi LPT Orsay  01-91\par
%\noi October 2001

\newpage
\pagestyle{plain}
\baselineskip=18 pt

\section{Introduction}

The NA50 interpretation of the data on
$J/\psi$ suppression in $pA$, $SU$ and $Pb$ $Pb$ collisions is as
follows \cite{1r}. The $pA$, $SU$ and peripheral $Pb$ $Pb$ data (up to $E_T \sim
35 \div 40$~GeV), are well described with nuclear absorption alone with
an absorptive cross-section $\sigma_{abs} = 6.4 \pm 0.8$~mb. At $E_T
\sim 40$~GeV there is a sudden onset of anomalous suppression. A second
accident occurs at $E_T \sim 90 \div 100$~GeV, close to the knee of
the $E_T$ distribution, where a change of curvature in the shape of the
suppression is observed -- followed by a steep fall-off. However, at
variance with this view, the
most peripheral point in $Pb$ $Pb$ has consistently lied above the NA50
nuclear absorption curve -- which extrapolates $pA$ and $SU$ data.
This tendency is now confirmed by the
preliminary data on the $J/\psi$ suppression versus the energy,
$E_{ZDC}$, of the zero degree calorimeter. Here several peripheral
points lie above the NA50 nuclear absorption curve and, 
more important, they exhibit a
steeper suppression pattern. It is, therefore,
important to study these data and to compare them with the results of
deconfining and comover interactions 
models\footnote{For reviews of deconfining and
comover interaction models, see
\protect{\cite{4r}}. Alternative models have also been proposed
\protect{\cite{5r}}.}. This comparison is also
of interest in the large $E_T$ region. Indeed,
the
NA50 data for $Pb$ $Pb$ collisions at large $E_T$ can be described
either in a deconfining
model \cite{2r} or as a result of the interaction with comovers
\cite{3r}. \par

   Our purpose here is to confront with each
other the results of two specific models, namely a deconfining
\cite{2r} and a comovers interaction \cite{3r} model -- and to compare
them with experiment.
In the
first model \cite{2r}, one requires either two sharp thresholds, as in
\cite{6r}, or a single threshold smoothed with an arbitrary function.
Both in this model and in the comovers one \cite{3r}, it is
necessary to introduce the fluctuations in $E_T$ \cite{7r} in order to
reproduce the fall-off of the data at large $E_T$ (beyond the ``knee''
of the $E_T$ distribution). It has been shown in \cite{2r} that the
effect of the $E_T$ fluctuations is significantly larger in the
deconfining scenario than in the comovers approach. This leads to
agreement with the large $E_T$ data in the first case -- while the
fall-off obtained in the comovers approach is too weak \cite{3r,7r}.
However, all data beyond the knee have been obtained in the so-called
Minimum Bias ($MB$) analysis, where only the ratio of $J/\psi$ to 
$MB$ cross-sections is
measured and it is divided by a theoretical ratio of $DY$ to $MB$ 
cross-sections, i.e.

\beq
\label{1e}
\left . R_{MB} = \left ( {J/\psi \over DY} \right )_{MB} = \left (
{J/\psi \over MB} \right )_{exper.}\right / \left ( {DY \over MB}\right
)_{th} \quad . \eeq

\noi In the theoretical model used by the NA50 collaboration \cite{8r},
the ratio $(DY/MB)_{th}$ saturates at large $E_T$. In contrast, it has been
argued in \cite{3r} that this ratio also falls at large $E_T$. This is
due to the shift in $E_T$ between the $DY$ and $MB$ event samples,
resulting from the $E_T$ taken by the dimuon. This shift is very small
(of the order of a few per mil) and, therefore, has no visible effect
up to the knee of the $E_T$ distribution -- where this distribution is
rather flat. However, it does have a sizeable effect in the tail, due
to the very steep fall-off of the distribution. Actually,
experimental data on the ratio $DY/MB$ are available \cite{8r}. They
do show a fall-off at large $E_T$, but the statistical errors are
large. Obviously, the same
effect should be present in the ratio $J/\psi$ over $MB$ and would 
cancel out in
the true ratio of $J/\psi$ to $DY$ cross-sections -- i.e. the one 
obtained in the
so-called standard analysis. However, it does not cancel in $R_{MB}$
because this effect is not included in the theoretical model for the
last factor in the r.h.s. of Eq. (\ref{1e}), used by the NA50
collaboration \cite{8r}. When this effect, as estimated in \cite{3r},
is taken into account, the comovers approach does describe
the data for $R_{MB}$, whereas, in the deconfining model \cite{2r}, the
fall-off beyond the knee is too strong. 
This is shown in Fig.~1 where the dashed lines are the results in the
comovers model \cite{3r} with and 
without $E_{T}$ loss. The upper solid line
is the result obtained in the 
deconfining model \cite{2r}. In this model one
calculates the true ratio of $J/\psi$ to $DY$ differential
cross-sections -- not $R_{MB}$. Therefore, this line should be
compared with the results in the standard analysis. To compare
with the data in Fig. 1, which where obtained with the $MB$ analysis,
one should take into account the $E_{T}$ loss. Incorporating
this effect, following the prescription \cite{3r}, one obtains the lower
full curve in Fig.~1.\par  

Clearly, a good way to avoid this problem is to measure $R_{MB}$
as a function of $E_{ZDC}$. Indeed, the main contribution to $E_{ZDC}$
is due to the energy of the spectator nucleons -- which is not affected by
the presence or absence of the dimuon trigger. \par

The plan of this paper is as follows. In Section 2 we derive the 
$E_{T}$ - $E_{ZDC}$ correlation. It turns out that this correlation
allows to relate to each other the $MB$ experimental distributions
in the two variables -- $E_{T}$ and $E_{ZDC}$. In Section 3 we
compute the $J/\psi$ suppression versus $E_{ZDC}$ and compare it
with the one measured in $E_{T}$, and with the results of the
deconfining model \cite{2r}. In Section 4, we compute the $J/\psi$
suppression in $pA$ and $SU$ in the comovers approach and compare 
the normalization in these systems with the one in $PbPb$.
The scenario for $J/\psi$ suppression resulting from the
comovers model is described in the Conclusions and confronted
with the deconfining one. \\

\section{\bf E$_{\bf T}$ - E$_{\bf ZDC}$ correlation.} 

\noi
In the models
\cite{2r}, \cite{3r}, one determines
the $J/\psi$ suppression as a function of the impact parameter, $b$.
However, $b$ is
not measurable and the NA50 collaboration uses, as a measure of
centrality, either $E_T$ or $E_{ZDC}$. It is, therefore, necessary to
determine both quantities as a function of $b$. Since $E_T$ is
proportional to the multiplicity, the relation between $E_T$ and $b$
results from the determination of the multiplicity at each $b$ in the
rapidity region of the $E_T$ calorimeter. In the comovers model
\cite{3r} this is
done \cite{9r} in the Dual Parton Model (DPM). More precisely, we put~:

\beq \label{2e} E_T(b) = {1 \over 2} \ q\ N_{yca}^{co}(b) \quad . \eeq

\noi Here $N_{yca}^{co}$ is the charged multiplicity in the rapidity
region of the $E_T$ calorimeter. The factor $1/2$ is introduced
because only the energy of neutrals is measured by the calorimeter.
Thus the coefficient $q$ is close to the average energy per particle.
However, the difference between multiplicities of positive negatives
and neutrals as well as the efficiency of the $E_T$ calorimeter do
affect the value of $q$. This value can be determined from the position
of the ``knee'' of the $E_T$ distribution of $MB$ events measured by
the NA50 collaboration. We obtain $q = 0.62$~GeV \cite{3r} (see below).
\par

The energy of the zero degree calorimeter is given by

\beq \label{3e} E_{ZDC}(b) = [A - n_A(b)] E_{in} + \alpha \ n_A(b) \
E_{in} \quad . \eeq

\noi Here $A - n_A(b)$ is the number of spectator nucleons of $A$ and
$E_{in} = 158$~GeV is the beam energy. While the first term in the
r.h.s. of Eq. (\ref{3e}) gives the bulk of $E_{ZDC}$, the latter corresponds
to the contamination by secondaries emitted very forward
\cite{10r} -- assumed to be proportional to the number
of participants, $n_A(b)$. Here also the value of $\alpha$ can be precisely
determined from the position of the ``knee'' of the $E_{ZDC}$ distribution
of the $MB$ event sample measured by NA50 \cite{10r}. We obtain $\alpha
= 0.076$. \par

Eqs. (\ref{2e}) and (\ref{3e}) give the relation between 
$b$ and $E_T$ and $b$ and $E_{ZDC}$, respectively. These
relations refer to average values and do not contain any
information about the tails of the $E_T$ or $E_{ZDC}$ distributions.
Eqs. (\ref{2e}) and (\ref{3e}) also lead to a correlation between
(average values of) $E_T$ and $E_{ZDC}$. This correlation is shown in
Fig.~2, and gives a good description of the experimental one
\cite{8r}. We see from Fig.~2 that the $E_T - E_{ZDC}$ correlation is
close to a straight line\footnote{This is due to the fact that
$N_{yca}^{co}(b)$ in Eq. (\protect{\ref{2e}}) is practically
proportional to $n_A(b)$ (see Fig.~1 of \protect{\cite{3r}}).} and
therefore can be accurately extrapolated beyond the knee of the $E_T$
and $E_{ZDC}$ distributions. It turns out that this extrapolation describes
the data quite well\footnote{One can understand the physical origin of
this extrapolation if one assumes that a fluctuation in $E_T$ is
essentially due to a fluctuation in $n_A$ -- which, in turn, produces a
corresponding fluctuation in $E_{ZDC}$, via Eq. (\protect{\ref{3e}}).}.
Let us discuss this point in detail. It is well known that the
correlation $E_T - b$ can be described by a Gaussian~:

\beq
\label{4e}
P(E_T, b) = {1 \over \sqrt{2\pi q a E_T(b)}} \exp \left [ - {[E_T -
E_T(b)]^2 \over 2qaE_T(b)} \right ]
\eeq

\noi with $q = 0.62$~GeV and $a = 0.60$ \cite{3r}. The resulting $MB$
distribution is indistinguishable from the solid line of \cite{10r}
(1998 data). The $MB$ distribution of the 1996 data \cite{10r} is reproduced
with $q = 0.62$ GeV and $a = 0.852$ \cite{3r}\footnote{
At first sight these sets of values look very different
from the ones used by the NA50 Collaboration.
Nevertheless, they reproduce the same $E_T$ distribution. This is due
to the fact that the product $qa$, which, according to eq.
(\ref{4e}), determines
the width of the distribution, is very similar in the two cases. As
for the difference in the values of $q$ it is just due to its
definition, which is different in the two
approaches (eq. (\ref{2e}), in our case).}.\par

Applying the $E_T - E_{ZDC}$ correlation resulting from Eqs. 
(\ref{2e}) and (\ref{3e}) we obtain the $E_{ZDC}$ distribution of 
$MB$ events shown in
Figs.~3. We see that the NA50 data are well described, not  only up to 
the knee, but also in the tail of the distribution. The turn-over in 
the data at
large $E_{ZDC}$ is due to the fact that they have not been corrected 
for efficiency.
The comparison of Figs.~3a and 3b is quite instructive. In
Fig.~3a it seems that the calculated curve has too much curvature
and overshoots the data at small $E_{ZDC}$. However, this is no
longer the case in Fig.~3b. Here, on the other hand, there is
some discrepancy in the region $6 < E_{ZDC} < 12$~TeV -- which
is not present in Fig.~3a. Finally, the calculated curve is slightly
too broad at the tail in Fig.~3a and too narrow in Fig.~3b. From
this comparison, we conclude that Eq. (4), supplemented with
the $E_{T} - E_{ZDC}$ correlation of Fig.~2, gives a good
description of the $E_{ZDC}$ distribution when both the 1996 and
the 1998 data are considered. \par

The consequences of this result are quite 
interesting. Indeed,
the $E_T - E_{ZDC}$ is essentially a correlation between multiplicity 
(which in the rapidity region of the calorimeter is practically 
proportional to
the number of participants) and number of spectators. Hence it cannot 
be affected by the dimuon trigger\footnote{Except in the tails,
where the $E_{T}$ loss affects the $E_{T}$ distributions of $J/\psi$
and $DY$ without changing the $E_{ZDC}$ ones.}. 
Therefore, not only 
the $MB$, but also
the $E_{ZDC}$ distributions of $J/\psi$ and $DY$ event samples 
can be obtained from the corresponding ones versus $E_T$ 
applying the $E_T -
E_{ZDC}$ correlation. This is an important result since it allows to 
relate to each other the $J/\psi$ suppression versus $E_T$ and 
versus
$E_{ZDC}$ (obtained with a different calorimeter) and check their 
consistency. \\

\section{J/$\psi$ suppression versus E$_{\bf ZDC}$.}

\noi  
We show in
Fig.~4 the results for the ratio of $J/\psi$ to $DY$ cross-sections versus
$E_{ZDC}$ obtained in the deconfining model \cite{2r} and in the comovers model
\cite{3r}. The curves, are obtained from the corresponding 
ones
versus $E_T$ (upper solid and dashed cuerves in Fig. 1), applying the $E_T - 
E_{ZDC}$ correlation given by
Eqs. (\ref{2e}) and (\ref{3e}) and shown in  Fig.~2. We keep the 
absolute normalizations unchanged. 
No $E_{T}$ loss is incorporated in these calculations (see below). \par

Comparing the data with the model predictions, we see that
a better description of the small $E_{ZDC}$ region is obtained
when the $E_T$ fluctuations are taken into account. This was to be
expected since the fluctuations in $E_T$ and $E_{ZDC}$ are related to
each other via the $E_T - E_{ZDC}$ correlation -- as shown in Fig.~2.
\par

As discussed above, the effect of the $E_T$ loss is not included here 
since $E_{ZDC}$ is not affected by the dimuon 
trigger\footnote{Actually, some effect of the $E_{T}$ loss could
be present in the $J/\psi$ suppression versus $E_{ZDC}$ via the
second term of Eq. (3). However, this effect should be rather small due
to the large rapidity separation of the dimuon trigger and the
zero degree calorimeter.}.
Nevertheless, the curve obtained in the comovers approach does 
describe the most central data while it failed to describe the 
data versus $E_T$ at large $E_T$ (see Fig. 1). This result lends support to 
the interpretation in ref. \cite{3r} -- namely 
that this
discrepancy is due to the $E_T$-loss induced by the $J/\psi$ trigger. \par

We also see in Fig.~4 that the fall-off for central events is stronger
in the deconfining than in the comovers model. The present data do
not allow to discriminate
between them. However, this should be possible when the shape and
absolute normalization of the data are better known - possibly with
the 2000 NA50 data.

Turning to the large $E_{ZDC}$ region (peripheral events), we see that
the data favor the comovers model. Indeed, it is clear that, in the
present data, the ratio $R_{MB}(E_{ZDC})$ at large $E_{ZDC}$ falls more
steeply than the NA50 nuclear absorption curve -- fitting $pA$ and 
$SU$ data. If confirmed, this would
clearly disfavor the deconfining model -- where nuclear absorption is
assumed to be the only
source of suppression below the deconfining threshold. On the contrary,
such a feature is expected in a comovers approach, and is clearly seen
in Fig.~4. 

An important observation from Fig. 4 is the fact that the
pattern of $J/\psi$ suppression in the NA50 data is different in
$E_T$ and $E_{ZDC}$ --
at least in the region $9 \ \lsim E_{ZDC} \ \lsim\ 14$ TeV. This is at
variance with the results in Section 2, according to which the
shapes of the two curves should be the same. More precisely, if the
shape of the $J/\psi$ suppression in $E_{ZDC}$ is different from the
one in $E_T$ for whatever reason (for instance, due to the different
calorimeters), the shape of the $MB$ distribution in the two variables
should also be different -- and it is not. This point needs to be
clarified.
It should be noted here that the ratio of $J/\psi$ to $DY$ differential
cross-sections obtained in the standard
analysis \cite{10r}, although
consistent with the one obtained in the $MB$ analysis, do not
show any sign of a different pattern with respect to the one versus
$E_{T}$. This is illustrated in Fig.~5 where the standard analysis
data are compared with the same theoretical curves of Fig.~4.
Note also that in the $MB$ analysis the absolute normalization
is not measured. It is determined from the one obtained with the standard
analysis. This determination can not
be precise
due to the uncertainty in
the shape of the $J/\psi$
suppression
discussed above.
Actually, the
adjustment of the absolute normalization has been done \cite{10r} in
the region $8 < E_{ZCD} < 17$~TeV -- where the difference in the
shape of the $E_{T}$ and $E_{ZDC}$ distributions is the largest. \par

Further insight in the comparison of comovers and deconfining 
models can be gained by studying the $J/\psi$ suppression in
lighter systems.\\ 

\section{J/$\psi$ suppression in pA and SU.}

\noi 
Let us 
compute the ratio $J/\psi$ over $DY$ in $SU$ at 200 GeV/c per nucleon 
in
the model \cite{3r}. We use, of course, the same  values of the 
parameters as in $PbPb$~: $\sigma_{abs} = 4.5$~mb 
and $\sigma_{co} = 1$~mb. To get this ratio versus
$b$, the only new ingredient is the multiplicity of comovers -- which 
is again computed in DPM, in the way described in \cite{9r}. In this case,
only $R(E_T)$ is available ($R(E_{ZDC})$ has not been measured in
$SU$). In $SU$, data do not extend beyond the knee of the
$E_T$-distribution. Therefore, effects such as $E_T$ fluctuations or 
$E_T$ loss are
not relevant here. To compute $R(E_T)$, we also need the $E_T - b$
correlation which is
parametrized as in Eq. (\ref{4e}). The parameters $q$ and $a$ have
been obtained from a fit of the $E_T$-distribution of $DY$ given in
\cite{11r}. We obtain $q = 0.69$~GeV and $a = 1.6$. \par

Our results are shown in Fig.~6a. We see that the $E_T$ dependence of
the suppression is reproduced -- in spite of the fact that our
$\sigma_{abs}$ is smaller than in the NA50 nuclear absorption model.
This is due to the suppression by comovers. Actually, the effect of
the comovers in $SU$ is comparatively small for
all values of $E_T$. However, its contribution increases with $E_T$
and compensates for the difference in the values of $\sigma_{abs}$ in
the two approaches. \par

A recent reanalysis of the $pA$ data at 450 GeV \cite{mor} leads to a value
$\sigma_{abs} = 4.7 \pm 0.8$~mb in good agreement with the one used in the comovers
approach. This value is  significantly lower that the one,
$\sigma_{abs} = 6.4$~mb, used in the
NA50 nuclear absorption curve. A value $\sigma_{abs} = 4$~mb is used
in the deconfining model of ref. \cite{12rnew}.
However, within the
deconfining approach such a value of $\sigma_{abs}$ has the drawback
of producing a
$J/\psi$ suppression pattern in $SU$ which is flatter
than the data -- unless, of course, one reduces substantially the
value of the
threshold, thereby introducing some anomalous suppression in SU. \par

Let us now discuss the absolute normalization of the curve in Fig.~6a
which is 46.8. This number has to be compared with the ratio of $J/\psi$
to $DY$ $pp$ cross-sections, rescaled at 200 GeV, which is $46.6 \pm 5$
\cite{12r}. This nice agreement between $pp$ and $SU$ tends to indicate
that the $pA$ data will also be reproduced (see also \cite{13r}). This
is indeed the case, as shown in Fig.~6b.

Our result gives an $A$-dependence between $pp$ and $pU$
$A^{\alpha}$ with $\alpha = 0.943$. This is to be compared with
the NA38 value $\alpha = 0.919 \pm 0.015$ \cite{12r} and with the E866
\cite{14r} one $\alpha = 0.955 \pm 0.02 \pm 1$~\% systematics.\par
 
We can now compare the absolute normalization of our curve for $pp$,
$pA$ and $SU$ (46.8) with the corresponding one for $Pb$ $Pb$ (59.4).
As we see, there is a 27~\% difference. According to the NA50 estimates
\cite{15r}, 9~\% of this difference is due to the rescaling from $SU$
at 200 GeV to $Pb$ $Pb$ at 158 GeV. This factor 1.09 contains both
energy and isospin corrections\footnote{N. Armesto (private
communication) has recalculated these corrections. He has confirmed the
NA50 results concerning isospin. However, he finds an energy dependence
of the $DY$ practically identical to the one of $J/\psi$ estimated by
NA50 -- leading to a ratio of $J/\psi$ to $DY$ cross-sections 
practically energy
independent. This would solve the problem with the rescaling in energy
used by NA50 in which the central value for the ratio $J/\psi$ over
$DY$ in $pp$ decreases between 158 GeV (48.9) and 200 GeV (46.6) and
increases from 200 to 450 GeV (54.7).}. Therefore, it remains a real
discrepancy of about 17~\% in the relative normalization of $pp$, $pA$
and $SU$, on the one hand, and $Pb$ $Pb$ on the other hand. This 
point needs clarification. \\

\section{Conclusions}

The deconfining -- NA50 scenario, has been described in
the first lines of this work.
The alternative scenario obtained in the comovers
model \cite{3r} is summarized in Fig.~7. The
curves in Fig.~7 are obtained from 
the ones in Figs.~1 ($PbPb$), 6a ($SU$)
and 6b ($pp$ and $pA$),
using the relation between $A$ and $L$ (in $pA$) and
$E_{T}$ and $L$
(in $SU$ and $PbPb$) given by NA38-NA50.
In this Figure we see that there are two different curves
for $pA$ (dashed line) and $SU$ (full line). This is due
to the effect of the comovers. Contrary to nuclear absorption,
which is a universal function of $L$, the comovers contribution
at a given value of $L$ is different in different systems.
In particular, its effect turns out to be negligible 
in $pA$. As already seen in Fig.~6a and 6b the agreement with
the data is good, both in $pA$ and $SU$. The $PbPb$ data are
not shown in this Figure. They follow quite well the shape of 
the theoretical curve (dashed line), as demonstrated in 
Fig.~1 (lower dashed line). However, their absolute 
normalization is 17~\% higher than the theoretical one,
as discussed above.
In the comovers model, the $J/\psi$ suppression in
$PbPb$ is
larger than the one in $SU$ (at the same $L$)
and significantly steeper - indicating an important
anomalous suppression, present even in the most peripheral
events. The data do 
support the latter result --
which favors the comovers scenario. For central events, it lends support to
the interpretation of ref. \cite{3r}, according to which the 
discrepancy between the comovers model and $R_{MB}$, eq. (\ref{1e}), 
is due to the $E_T$
loss induced by the $J/\psi$ trigger. Indeed, the data versus 
$E_{ZDC}$ (which are not affected by this $E_T$ loss) are well
reproduced. Hopefully, a clear-cut discrimination between the 
deconfining and comovers scenarios will be possible with the 2000 
NA50 data. \\

\noi {\Large \bf Acknowledgments} \\

We thank N. Armesto, E. G. Ferreiro, A. Kaidalov and C. A. Salgado
for discussions and E. Scomparini for information
on the NA50 data. D. S. thanks Fundaci\'on Barrie de la Maza for
financial support. This work was supported in part by the European
Community-Access to Research Infrastructure Action of the Improving
Human Potential Program.

\newpage \section*{Figure Captions}

\noi {\bf Fig. 1 :} Ratio of $J/\psi$ to $DY$ cross-sections versus
$E_{T}$ in $Pb$ $Pb$ collisions at 158 GeV per nucleon 
in the deconfining model \cite{2r} (solid) and in the
comovers model \cite{3r} (dashed). In both cases the upper (lower) curves are 
obtained without (with) $E_{T}$ loss.\\ 

\noi {\bf Fig. 2 :} The correlation $E_T - E_{ZDC}$ obtained from
Eqs. (\ref{2e}) and
(\ref{3e}) and its extrapolation to the tails of these distributions
(see main text).\\

\noi {\bf Fig. 3 :} $E_{ZDC}$ distribution of $MB$ events obtained from
Eq. (\ref{4e}) applying the $E_T - E_{ZDC}$ correlation of Fig.~2. The
data in Fig. 3a were taken in 1998 \cite{10r} and those in Fig. 3b
were taken in 1996 \cite{10r}.\\ 

\noi {\bf Fig. 4 :}  Ratio of $J/\psi$ to $DY$ cross-sections versus $E_{ZDC}$
in $Pb$ $Pb$ collisions at 158 GeV per nucleon in the deconfining model
\cite{2r} (solid) and in the comovers model \cite{3r} (dashed). These
curves are obtained from the corresponding ones versus $E_T$ 
(upper solid and dashed lines in Fig. 1)
applying the $E_T - E_{ZDC}$ correlation of Fig.~2. The
dotted line is obtained without $E_T$ fluctuations. The data, 
obtained with the minimum bias analysis, are from ref. \cite{10r}.
The NA50 nuclear absorption curve \cite{10r} is also
shown.\\

\noi {\bf Fig. 5 :} Same theoretical curves as in Fig. 4,
compared with the data obtained in the standard analysis \cite{10r}.\\ 

\noi {\bf Fig. 6a :} The ratio of $J/\psi$ to $DY$ cross-sections as 
a function of $E_T$ in
$SU$ collisions at 200 GeV per nucleon obtained in the
comovers model \cite{3r}.
The data are from \cite{11r}.\par
 
\noindent {\bf 6b :} The ratio $J/\psi$ over $A$ in $pp$ and
$pA$ collisions at 450 GeV as a function of $A$
obtained in the comovers model \cite{3r}. The data are from
\cite{12r}. \\

\noi {\bf Fig. 7 :} The ratio of $J/\psi$ to $DY$ cross-section versus $L$
\cite{8r} for $pp$, $pA$ (dotted line) and $SU$ at 200 GeV
(solid line) and $PbPb$ (dashed line)
at 158 GeV, obtained in the
comovers model \cite{3r}
with a common normalization 46.8. Note that this normalization
is not calculable in the model. The data are from \cite{11r} \cite{12r}.\\

\newpage

\centerline{\bf Fig. 1}
\begin{figure}[hbtp]
\begin{center}
\mbox{\epsfig{file=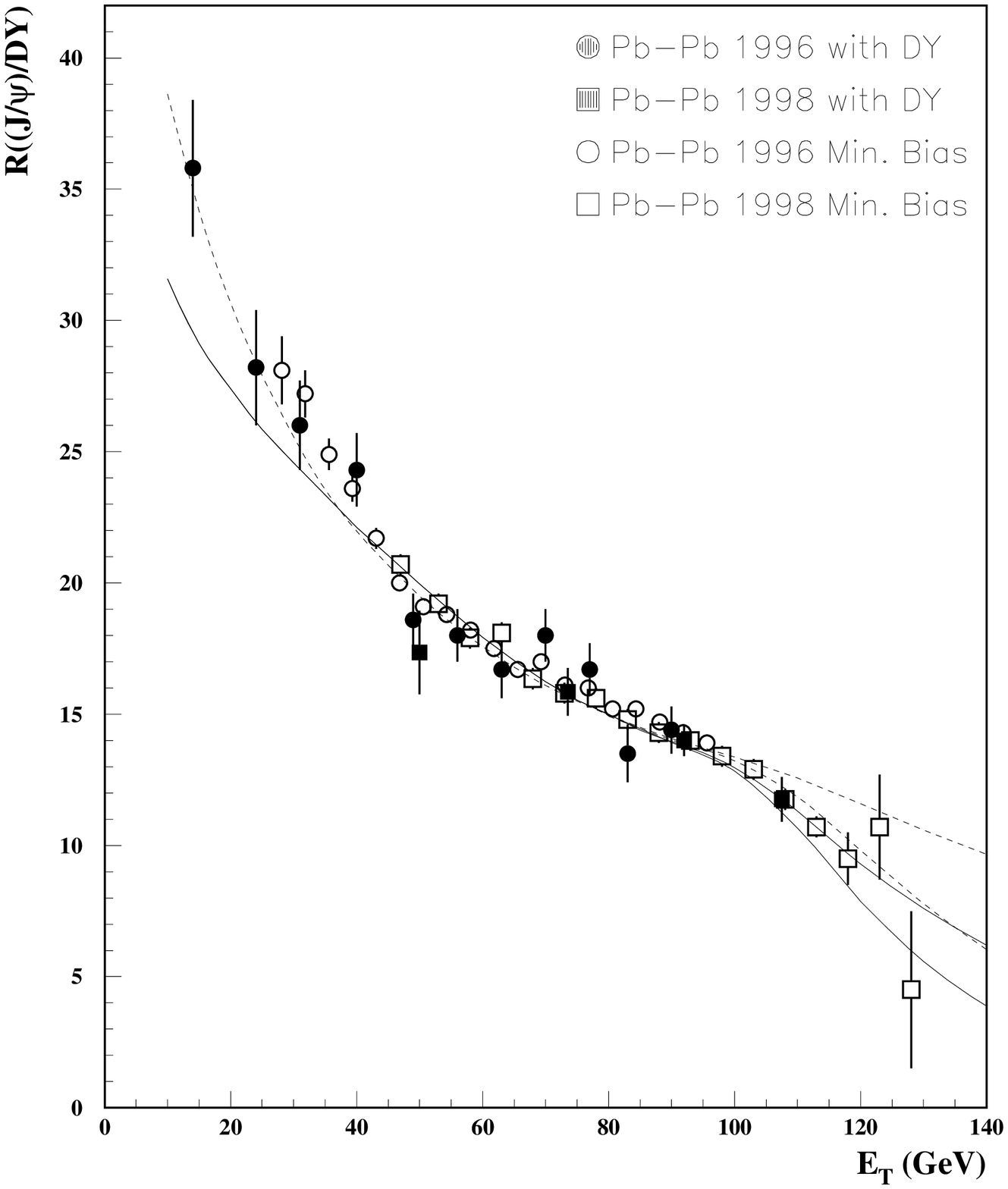,height=16cm}}
\end{center}
\end{figure}

\newpage
\centerline{\bf Fig. 2}
\begin{figure}[hbtp]
\begin{center}
\mbox{\epsfig{file=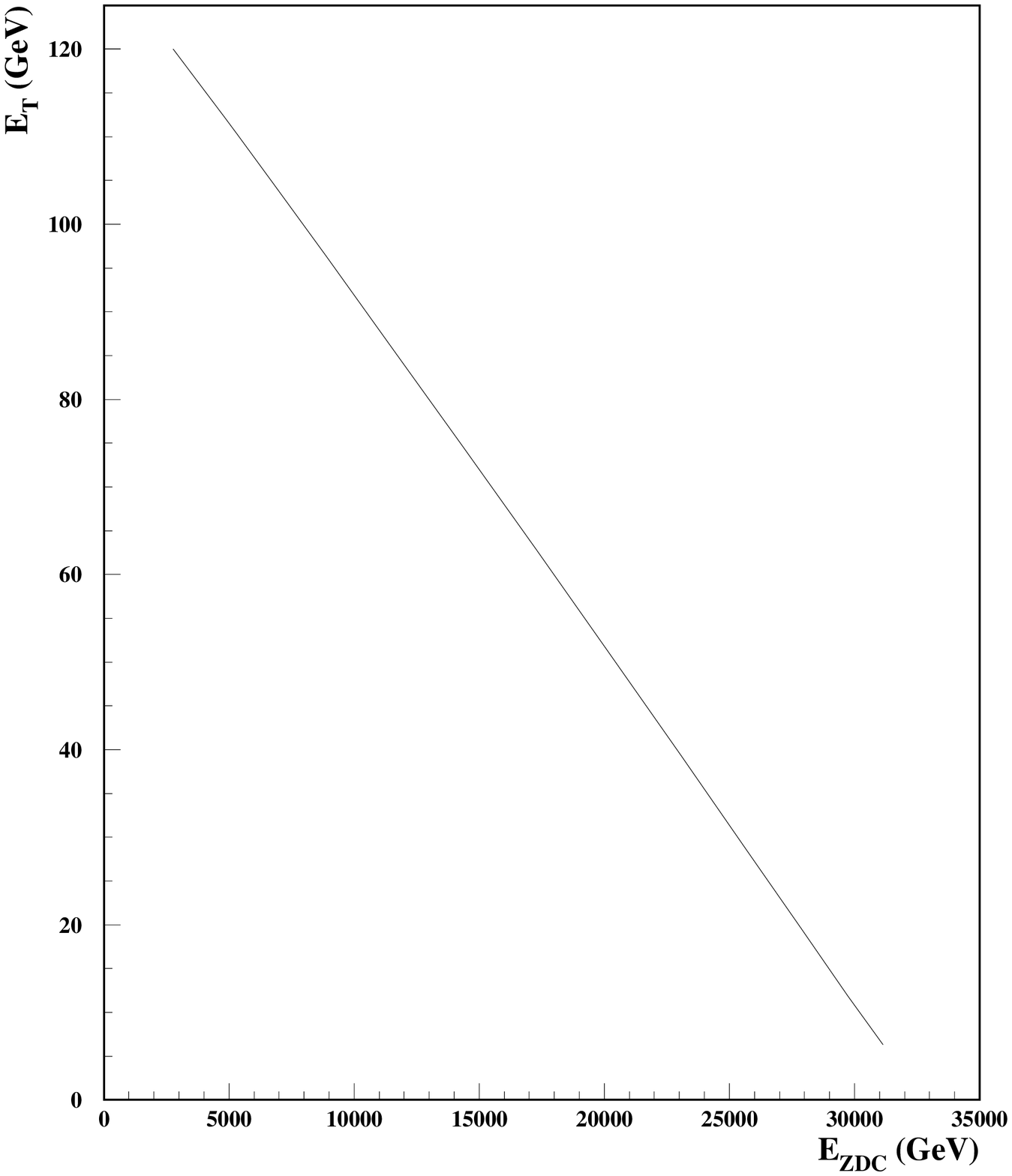,height=16cm}}
\end{center}
\end{figure}

\newpage
\centerline{\bf Fig. 3a}
\begin{figure}[hbtp]
\begin{center}
\mbox{\epsfig{file=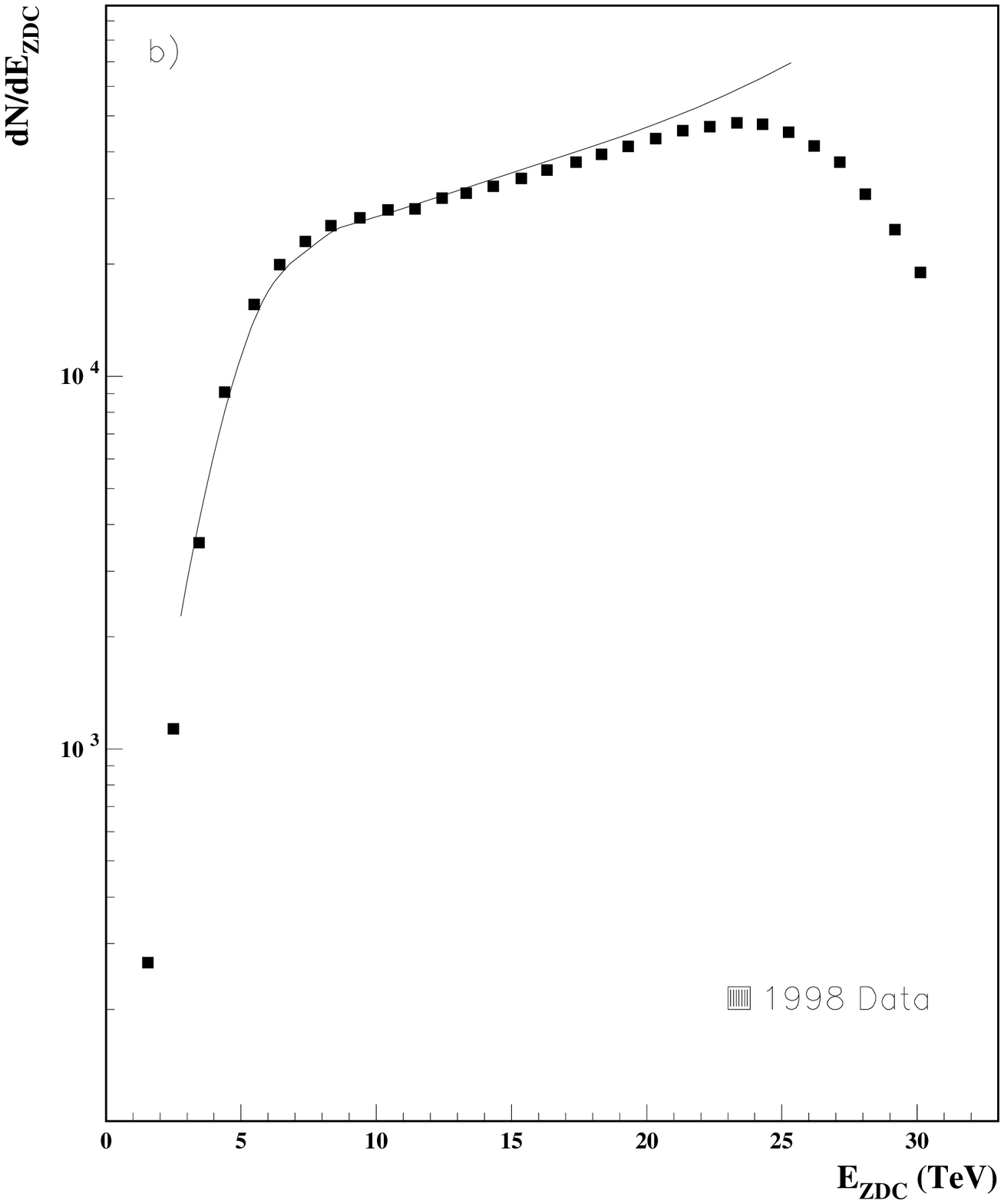,height=16cm}}
\end{center}
\end{figure}

\newpage
\centerline{\bf Fig. 3b}
\begin{figure}[hbtp]
\begin{center}
\mbox{\epsfig{file=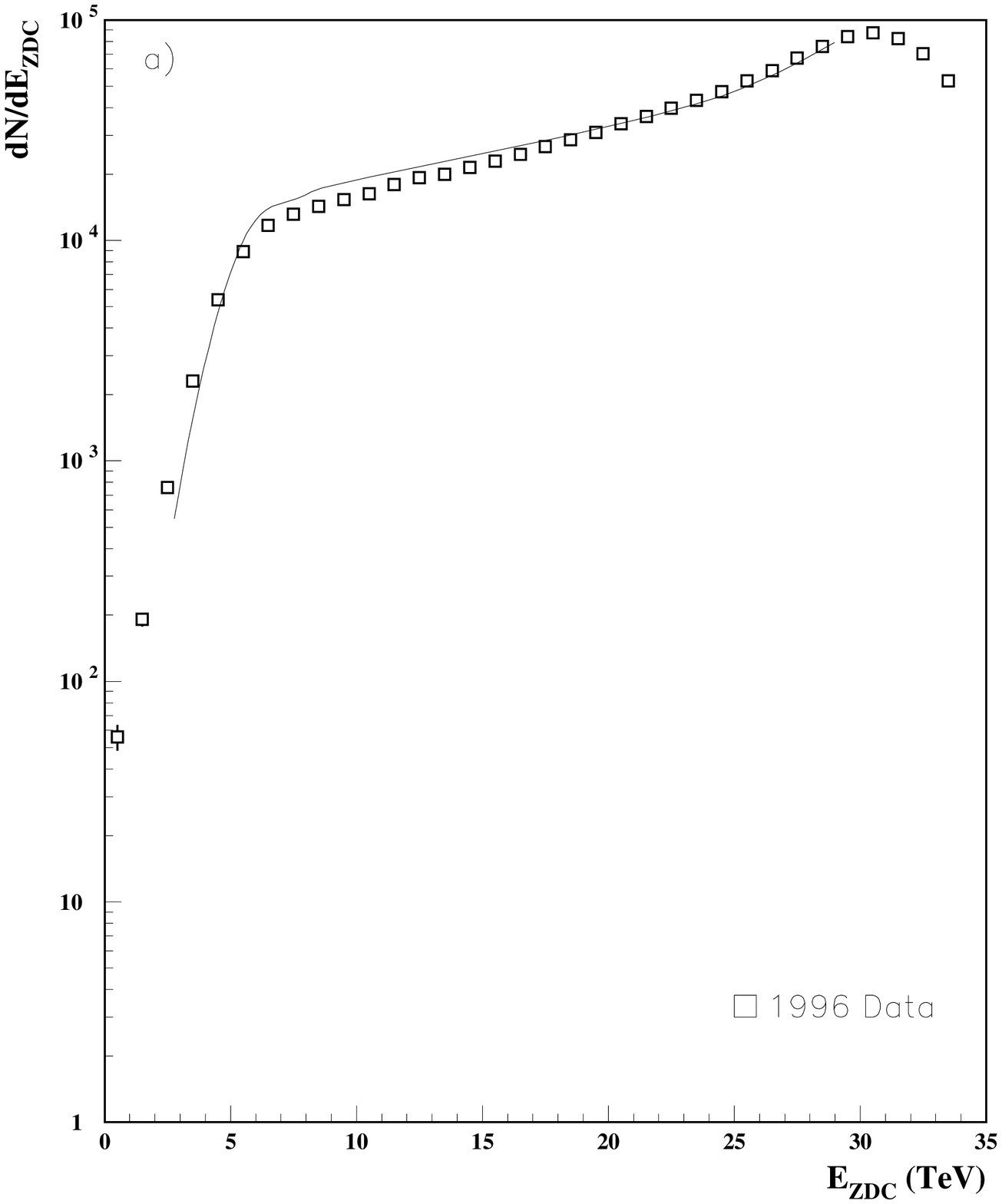,height=16cm}}
\end{center}
\end{figure}

\newpage
\centerline{\bf Fig. 4}
\begin{figure}[hbtp]
\begin{center}
\mbox{\epsfig{file=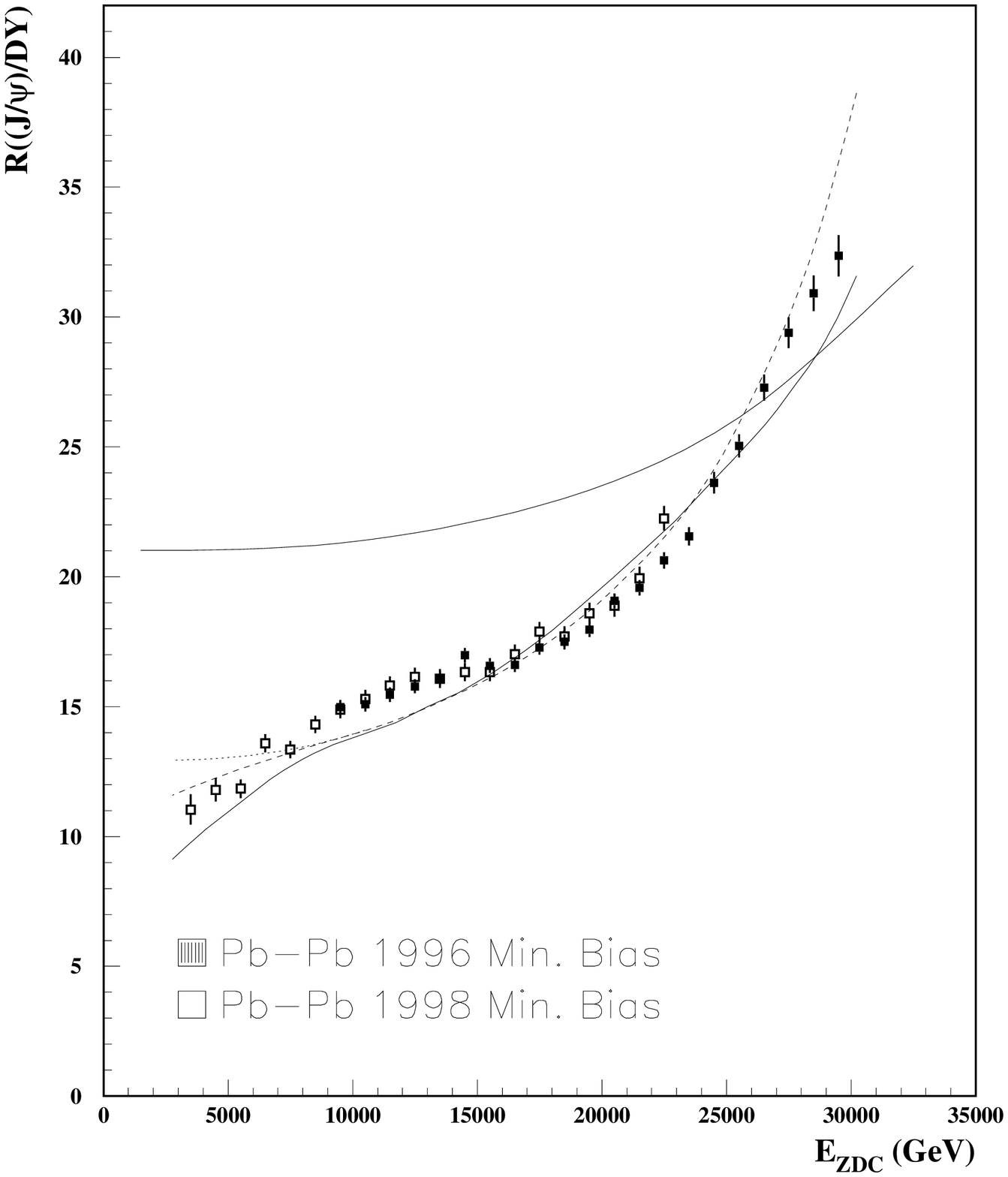,height=16cm}}
\end{center}
\end{figure}

\newpage
\centerline{\bf Fig. 5}
\begin{figure}[hbtp]
\begin{center}
\mbox{\epsfig{file=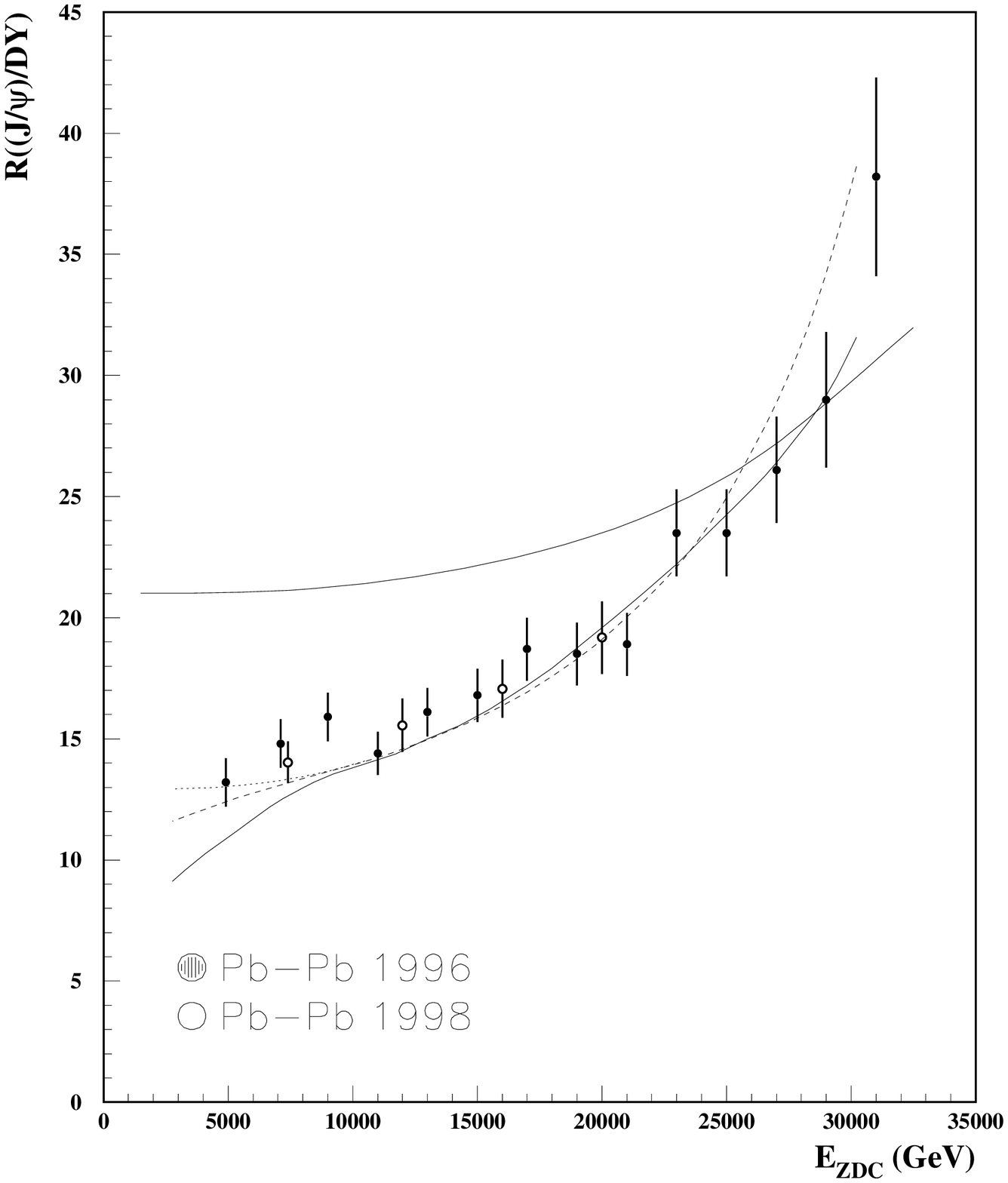,height=16cm}}
\end{center}
\end{figure}

\newpage
\centerline{\bf Fig. 6}
\begin{figure}[hbtp]
\begin{center}
\mbox{\epsfig{file=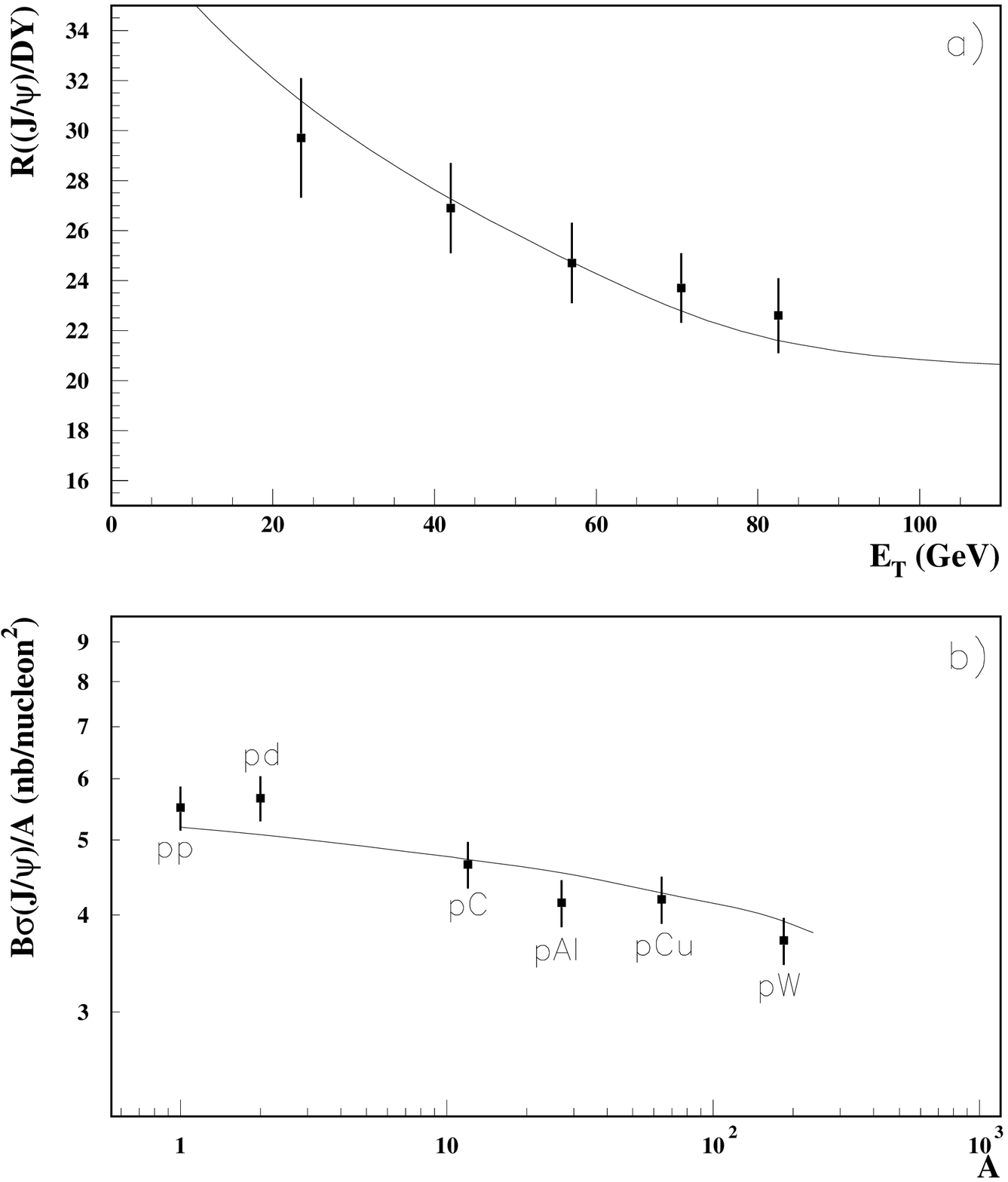,height=16cm}}
\end{center}
\end{figure}

\newpage
\centerline{\bf Fig. 7}
\begin{figure}[hbtp]
\begin{center}
\mbox{\epsfig{file=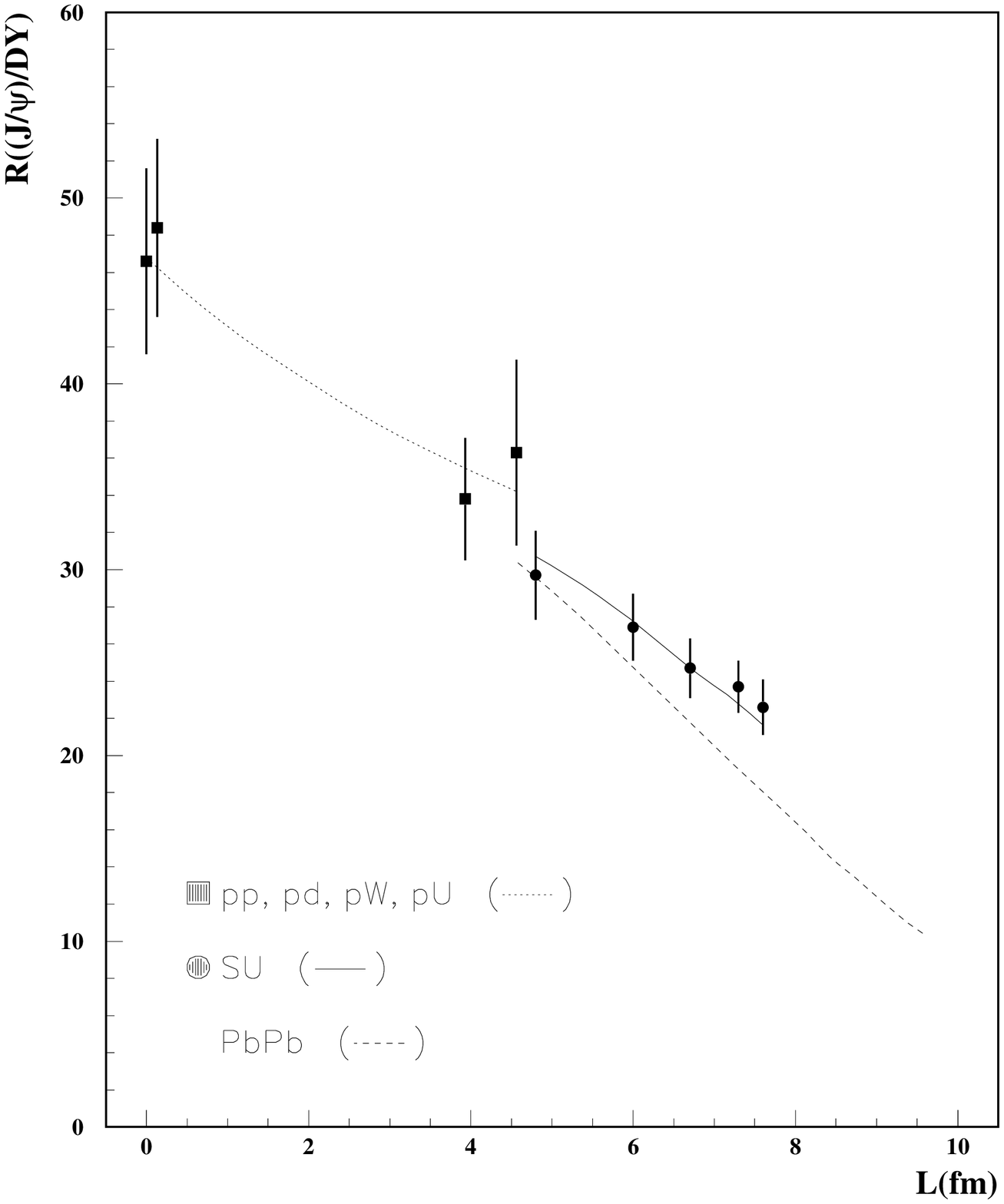,height=16cm}}
\end{center}
\end{figure}

\end{document}